\documentclass[psfig]{llncs}

\usepackage{multicol} 
\usepackage{makeidx}  
\usepackage{psfig}

\begin{document}
\title{Disorder Effects in CA-Models for Traffic Flow}
\titlerunning{Disorder Effects}  
\author{W. Knospe\inst{1}, L. Santen\inst{2},
  A. Schadschneider\inst{2}, M. Schreckenberg\inst{1}}

\institute{Theoretische Physik FB10\\
  Gerhard-Mercator Universit\"at Duisburg \\
  D-47048 Duisburg, Germany \\
  $ $ 
  \and Institut f\"ur Theoretische Physik\\ 
  Universit\"at zu K\"oln \\ D-50937 K\"oln, Germany}
%
%
%
\maketitle              

\begin{abstract}
We investigate the effect of quenched disorder in the
Nagel-Schreckenberg model of traffic flow. Spatial inhomogenities, 
i.e. lattice sites where the braking probability is enlarged, are
considered as well as particle disorder, i.e. cars of a 
different maximum  velocity. Both types of disorder lead to segregated
states. 
\end{abstract}

\section{Introduction}

The formation of traffic jams is one of the fundamental problems of
traffic flow theory. Traffic jams can form spontanously 
as well as due to hindrances, e.g. road works or slow
cars. Although these hindrances often cover only a small part of
the road, they can cause large jams, or, in a more physical language, one
observes macroscopic effects due to local defects in the system. This
behaviour is one of the characteristic properties of nonequilibrium
systems. In the context of driven lattice gases one can distinguish in
principle two types of defects: 1) lattice
defects, e. g. sites where the mobility of particles is reduced, and
2) particle defects, for example 'slow' particles.   

For the simplest model for a driven lattice gas in one dimension, 
the asymmetric exclusions process (ASEP)\cite{nr}, it has been shown that both
types of defects can cause phase transitions. Slow
particles, i.e. particles with a reduced hopping
rate, determine completely the flow at small densities. Moreover, if
the distribution of the hopping rates fulfills certain conditions, one
can show analytically that the variance of the distance distribution
has a logarithmic divergence at the transition density
\cite{Ferrari,Krug,Mallick,Evans}.

Also lattice defects in the ASEP have been investigated
extensively
\cite{Janowski,Lebowitz,Tripathy,Emmerich,Yukawa,ernst}. Implementing 
lattice defects one can 
observe three different phases: A high and a low density phase, where
the average flow of the homogeneous system is recovered, and a
segregated phase at intermediate densities, where the flow takes a
 constant value. Despite extensive effort, exact analytical results
 exist only for a special type of update \cite{hinrichsen}, but approximative
 descriptions are in  reasonable agreement with simulation
 results. 

In this work we generalize the results of the ASEP to the
Nagel-Schreckenberg (NaSch) model of traffic flow (for a detailed
explanation of the model see \cite{NaSch,duiproc}). Compared with the
ASEP, the particles (cars) can hop more than one site in a single
update step and the update rules are applied in parallel to all cars. 
Therefore slow cars can be considered in two ways. First, one
can think of cars with a smaller maximum velocity and second, of cars with an
enlarged braking probability. This has been done very recently and
analogous results to the $v_{max}=1$ case have been found \cite{Chowdhury}.
In contrast to \cite{Chowdhury} we discuss the case of two different maximum
velocities in the third chapter.
In the next chapter we show simulation results for the
NaSch model with defect sites, implemented as sites where the braking
probability of cars is higher compared to the rest of the lattice. An
alternative choice has been used in \cite{Csahok}, where a 'speed
limit' in a part of the lattice has been considered.

\section{Defect Sites}

As already mentioned above, in this chapter we show simulation results of
the NaSch model on a lattice with defect sites.      
\begin{figure}[h]
 \centerline{\psfig{figure=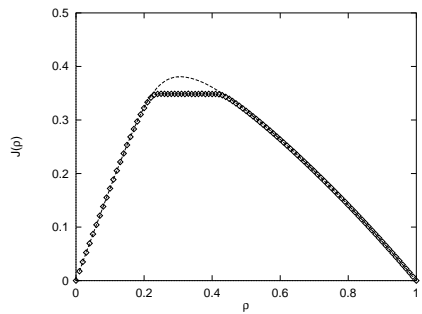,bbllx=50pt,bblly=50pt,bburx=550pt,bbury=400pt,height=4cm}}
\caption{\protect{Fundamental diagram of the NaSch model with a
    single defect site. The model parameters are given by $v_{max}=2$
    (maximum velocity), $p=0.25$ (braking probability), $p_d = 0.75$
    (braking probability at the defect site)  and $L=3200$ (system size).}}
\label{single_fund}
\end{figure}
Fig. \ref{single_fund} shows the fundamental diagram of a system with
a {\em single} defect site. Obviously we can distinguish three different
phases depending on the 
density. In the high and low density phases the average flow of the
defect systems takes the same value as in the homogeneous system. For
intermediate densities the flow is constant and limited by 
the capacity of the defect site.

\begin{figure}[h]
 \centerline{\psfig{figure=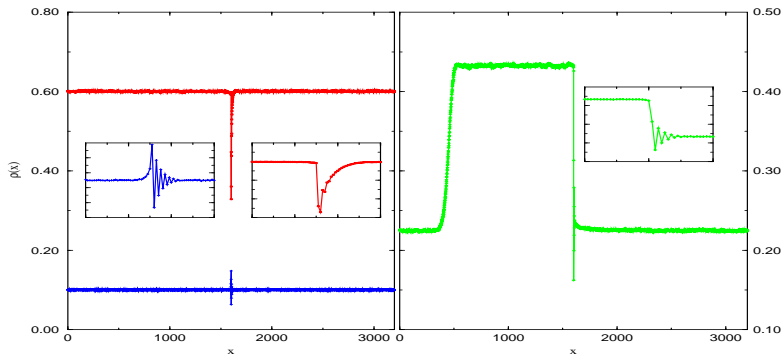,bbllx=0pt,bblly=0pt,bburx=1180pt,bbury=590pt,height=5cm}}
\caption{\protect{Density profiles in the three phases. In the high
    (red, $\rho =0.60$) and low (blue, $\rho =0.10$) density phases only
    local inhomogenities 
    occur near the defect site (the defect site is located at $x_d = 1600
    = L/2$), but for intermediate densities one observes phase
    separation (green, $\rho =0.30$). We used the same parameters as in
    Fig. \ref{single_fund}.}}
\label{single_prof}
\end{figure}

This behaviour of the average flow can be explained looking at the
density profile. In the high and low density phase only local
deviations from a constant profile can be observed, but at intermediate
densities one observes a separation into  macroscopic high and low density
regions. Changing the global density within the segregated phase, 
the bulk densities in the high ($\rho_h$)  and low density region
($\rho_l$) remain constant, only the length of the high 
and low density region changes. Consequently the average flow is constant in
the segregated phase, because the average density in the vicinity of
the defect site does not depend on the global density.  

Near the average position of the shock the density profile decays
exponentially from $\rho_h$ to $\rho_l$. Therefore one can
introduce a length scale $\xi$, which corresponds to the magnitude of the
fluctuations of the shock position. For $v_{max} >1$ we found
numerically that $\xi \sim \sqrt{L}$ for all densities in the
segregated phase  we took into account. This scaling behaviour is
already known for the ASEP at $\rho \neq 0.5$. The modified scaling
behaviour $\xi \sim L^{1/3}$ at $\rho = 0.5$ was not found for
$v_{max}>1$. This confirms the picture that the reduction of
fluctuations is a consequence of the particle-hole symmetry \cite{Janowski}.   

A good estimate for the plateau value of the flow for the case of
$v_{max}=1$ can be obtained
using the assumption that the system is separated into two regions of
constant density, where the results from the homogeneous system can be
used. Using an argumentation similar to \cite{Yukawa} we obtain  
the plateau value of the  average flow: 

\begin{equation}
  \label{flow_plateau}
  J_p = \frac{1}{2} \left( 1-\frac{1}{q+q_d}\sqrt{(q+q_d)^2 - 4q^2q_d} \right)
\end{equation}

The bulk value of the density in the high (low) density region the
density is given by $\rho_h = \frac{q}{q+q_d}$ ($\rho_l =
\frac{q_d}{q+q_d}$) with $q =1-p$ and $q_d =1-p_d$. 
The shock is located at $r = x_d - \frac{\rho-\rho_h}{\rho_h-\rho_l}L$
if $r\geq0$ or at $r'= L-r$ 
if $r<0$. This approximative treatment
of the defect system is in a resonable agreement with the simulation
results for large system sizes (for a more detailed discussion see
\cite{future}).  In principle one could
treat higher velocities in the same way, but unfortunately no exact
analytical description of the homogeneous system has been found by so
far.

\section{Particle defects}

In order to study the effect of slow cars on the throughput of single
lane traffic, we discuss a system which consists of fast cars with maximum
velocity $v_f=3$ {\em and} one slow car with $v_s=2$.

\begin{figure}[h]
 \centerline{\psfig{figure=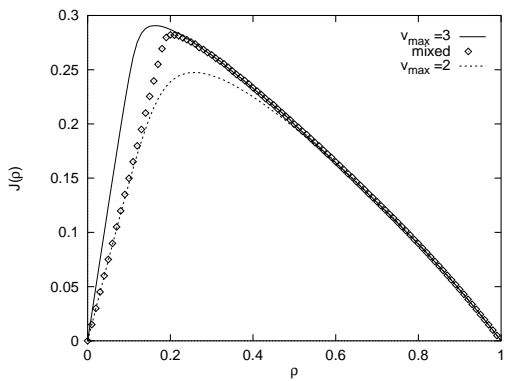,bbllx=50pt,bblly=50pt,bburx=550pt,bbury=410pt,height=5cm}}
\caption{\protect{Fundamental diagram of a system with one slow
    car (dots) compared with fundamental diagrams of the corresponding 
    homogeneous systems.  The
    maximum velocity of the slow car is given by $v_{s}=2$ and of the
    other cars by $v_{f}=3$. The braking probability of all cars is
    $p=0.5$. }}
\label{slcfund}
\end{figure}

In Fig. \ref{slcfund}  the fundamental diagram of this system is
compared with the fundamental diagram of the analogous homogeneous
systems. Obviously for low densities the flow is given by
$J_l(\rho) = \rho (v_{s}-p)$ in a agreement with the homogeneous
system with $v_{max} =2$. Compared to the homogeneous system one 
observes the linear density dependence also for larger values of the
average density. The average flow depends linearly on the global
density if $J_l(\rho) \leq J_{v_f}(\rho)$ holds, where  $J_{v_f}(\rho)$
denotes the stationary flow of the homogeneous system with fast cars.
Deviations from this form that are observable near the intersection
point of $J_l(\rho)$ and  $J_{v_f}(\rho)$, are due to the finite size
effects.

\begin{figure}[h]
 \centerline{\psfig{figure=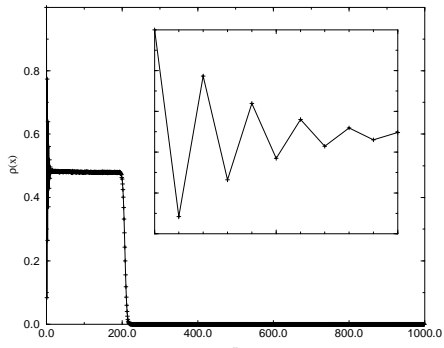,bbllx=50pt,bblly=50pt,bburx=650pt,bbury=550pt,height=5cm}}
\caption{\protect{Density profile relative to the position of the slow
    car. The data are obtained for the following set of parameters:
    $v_s =1$, $v_f=2$, $p=0.5$ and, $\rho =0.1$. }}
\label{slcprof}
\end{figure}

The reason for the stability of the free flow regime in the mixed
system is the gap in front of the slow car in the stationary state. In
Fig. \ref{slcprof}  the density profile  relative to the position of
the slow car is shown. Obviously  a large jam behind the slow car
appears and the density in front of the slow car vanishes. Therefore
the slow car can move with its free flow velocity $v_s-p$. This velocity
determines the velocity of the whole jam and therefore the average
velocity of the cars is simply given by $J_l(\rho)$. This is one
possible simple scenario for the occurance of 'moving' jams that are
well known from measurements. 

In addition to the macroscopic form, two details of the density profile
have to be discussed. Behind the slow car one observes an oscillating
amplitude of the density profile. These oscillations are well know
from the spatial correlation function of the homogeneous system for
small $p$ and densities near $\rho_c =1/(v_{max}+1)$ \cite{eisi}. Obviously the
slow car synchronizes the motion of the faster cars. Furthermore we
want to discuss the decay of the density profile at the end of the
jam. For finite values of the braking probability the jamlength is
fluctuating. These fluctuations are responsible for the finite size
effects discussed above, because temporarily the gap in front of the
slow car vanishes due to fluctuations of the jam length. Measurements
of the jamlength show that the fluctuations scale according to $\Delta
L_J \sim \sqrt{L}$.  Due to the subextensive scaling of the fluctations
the segregated states are observable up to $\rho = \rho_J$ if 
$L \rightarrow \infty$, where $\rho_J$ denotes the bulk density of the
jam. 

\section{Summary}

The investigation of local defects in the NaSch model has shown that
these defects can change the macroscopic properties of the model. In
the case of lattice defects three phases occur analogous to the ASEP 
(i.e. $v_{max} =1$) with continuous time update. The segregated phase
also exists for 
higher maximum velocities, but some details of the density profile are
changed. In the low density regime one observes an oscillating
profile as a consequence of the parallel update. The anomalous scaling
of the fluctuations of the shock position found for the ASEP at $\rho
= 0.5$ is absent for $v_{max}>1$, for all 
parameter combinations that have been taken into account. This result
confirms the picture that the particle-hole symmetry is essential for
the reduction of fluctuations. 

Particle defects also produce phase separated stationary states at
low densities. These states consist of a large jam behind the slowest
vehicle and a large gap in front of the slowest car. The velocity of
the 'moving' jam is completely determined by the free flow velocity of
the slow car. The formation of large clusters due to slow cars can
also be observed in two-lane traffic, but due to lane changing the
lifetime of the large clusters is finite (at least for low
densities).  Nevertheless it has been shown,
that already a small concentration of slow cars changes leads to a
drastic reduction of the average flow at low densities
\cite{wolf}. 

Our results show that the system poperties 
can change completely, if local disorder is taken into
account. These results might also be important for an understanding of
the behaviour of real traffic which are often determined by
'imperfections'. Such imperfections are not only the defects
considered here, but also other deviations from an ideal system, e.g. ramps or
a finite system size.

%
%
%

\end{document}